\documentclass[aps,prd,showpacs,amsmath,amssymb]{revtex4}
\usepackage{epsfig}
\begin{document}

\title{How to approach continuum physics in lattice Weinberg - Salam model}

\author{M.A.Zubkov}
 \email{zubkov@itep.ru}
\affiliation{ITEP, B.Cheremushkinskaya 25, Moscow, 117259, Russia}

\today

%date{July, 12, 2003}

\begin{abstract}
We investigate lattice Weinberg - Salam model without fermions  numerically for
the realistic choice of coupling constants correspondent to the value of the
Weinberg angle $\theta_W \sim 30^o$, and bare fine structure constant around
$\alpha \sim \frac{1}{150}$. We consider the values of the scalar self coupling
corresponding to Higgs mass $M_H \sim 100, 150, 270$ GeV. It has been found
that nonperturbative effects become important while approaching continuum
physics within the lattice model. When the ultraviolet cutoff $\Lambda =
\frac{\pi}{a}$ (where $a$ is the lattice spacing) is increased and achieves the
value around $1$ TeV one encounters the fluctuational region (on the phase
diagram of the lattice model), where the fluctuations of the scalar field
become strong. The classical Nambu monopole can be considered as an embryo of
the unphysical symmetric phase within the physical phase. In the fluctuational
region quantum Nambu monopoles are dense and, therefore, the use of the
perturbation expansion around trivial vacuum in this region is limited. Further
increase of the cutoff is accompanied by a transition to the region of the
phase diagram, where the scalar field  is not condensed (this happens at the
value of $\Lambda$ around $1.4$ TeV for the considered lattice sizes). Within
this region further increase of the cutoff is possible although we do not
observe this in details due to the strong fluctuations of the gauge boson
correlator. Both mentioned above regions look unphysical. Therefore we come to
the conclusion that the maximal value of the cutoff admitted within lattice
Electroweak theory cannot exceed the value of the order of $1$ TeV.
\end{abstract}


\maketitle

\section{Introduction}

 It is well - known \cite{M_W_T}, that the finite temperature
perturbation expansion breaks down at the temperatures above the electroweak
transition/crossover already for Higgs masses above about $60$ GeV. Therefore the present
lower bound on the Higgs mass requires the use of nonperturbative techniques while
investigating electroweak physics at high temperature.

Nambu monopoles are not described by means of a perturbation expansion around the trivial
vacuum background. Therefore, nonperturbative methods should be used in order to
investigate their physics. However, their mass is estimated at the Tev scale. That's why
at zero temperature and at the energies much less than $1$ Tev their effect on physical
observables is negligible. However, when energy of the processes approaches $1$ Tev we
expect these objects influence the dynamics. Recently the indications in favor of this
point of view were indeed found \cite{BVZ2007,VZ2008,Z2009}.

In this paper we consider lattice realization of zero temperature Electroweak
theory (without fermions). The phase diagram of the correspondent lattice model
contains physical Higgs phase, where scalar field is condensed and gauge bosons
$Z$ and $W$ acquire their masses. This physical phase is bounded by the phase
transition surface. Crossing this surface one leaves the Higgs phase and enters
the phase of the lattice theory, where the scalar field is not condensed.

In the lattice theory the ultraviolet cutoff is finite and is equal to the
momentum ${\Lambda} = \frac{\pi}{a}$ (see, for example,  \cite{UV}), where $a$
is the lattice spacing. The physical scale can be fixed, for example, using the
value of the $Z$-boson mass $M^{\rm phys}_Z \sim 90$ GeV. Therefore the lattice
spacing is evaluated to be $a \sim [90\,{\rm GeV}]^{-1} M_Z$, where $M_Z$ is
the $Z$ boson mass in lattice units. Within the physical phase of the theory
the lines of constant physics (LCP) are defined that correspond to constant
renormalized physical couplings (the fine structure constant $\alpha$, the
Weinberg angle $\theta_W$, and Higgs mass to Z-boson mass ratio $\eta =
M_H/M_Z$). The points on LCP are parametrized by the lattice spacing. Our
observation is that the LCP corresponding to realistic values of $\alpha$,
$\theta_W$, and $\eta$ crosses the transition between the two "phases" at a
certain value  $a = a_c$ and for $a < a_c$ the scalar field is not condensed.
We denote the corresponding value of the cutoff $\Lambda_c = \frac{\pi}{a_c}$.
Our estimate for the considered values of the Higgs mass $M_H \sim 100, 160,
270$ Gev is $\Lambda_c = 1.4 \pm 0.2$ Tev (for the considered lattice sizes).
We do not observe the dependence of $\Lambda_c$ on the lattice size.  That's
why the value $\Lambda_c$ might appear as the maximal possible value of the
cutoff allowed in the conventional Electroweak theory.

It is important to compare this result with the limitations on the Ultraviolet
Cutoff, that come from the perturbation theory. From the point of view of
perturbation theory the energy scale $1$ TeV appears in the Hierarchy problem
\cite{TEV}. Namely, the mass parameter $\mu^2$ for the scalar field receives a
quadratically divergent contribution in one loop. Therefore, the initial mass
parameter ($\mu^2= - \lambda_c v^2$, where $v$ is the vacuum average of the
scalar field) should be set to infinity in such a way that the renormalized
mass $\mu^2_R$ remains negative and finite. This is the content of the
so-called fine tuning. It is commonly believed that this fine tuning is not
natural \cite{TEV} and, therefore, one should set up the finite ultraviolet
cutoff $\Lambda$. From the requirement that the one-loop contribution to
$\mu^2$ is less than $10 |\mu_R^2|$ one derives that $\Lambda \sim 1$ TeV.
However, strictly speaking, the possibility that the mentioned fine tuning
takes place is not excluded.

In the perturbation theory there is also more solid limitation on the
Ultraviolet cutoff. It appears as a consequence of the triviality problem,
which is related to Landau pole in scalar field self coupling $\lambda$ and in
the fine structure constant $\alpha$. The Landau pole in fine structure
constant is related to the fermion loops and, therefore, has no direct
connection with our lattice result (we neglect dynamical fermions in our
consideration). Due to the Landau pole the renormalized $\lambda$ is zero, and
the only way to keep it equal to its measured value is to impose the limitation
on the cutoff. That's why the Electroweak theory is usually thought of as a
finite cutoff theory. For small Higgs masses (less than about $350$ GeV) the
correspondent energy scale $\Lambda_c^{0}$ calculated within the perturbation
theory is much larger, than $1$ Tev. In particular, for $M_H \sim 300$ GeV we
have $\Lambda_c^{0}\sim 1000$ TeV. It is worth mentioning that for $\lambda
\rightarrow \infty$ the perturbation expansion in $\lambda$ cannot be used. In
this case Higgs mass approaches its absolute upper bound\footnote{According to
the previous investigations of the $SU(2)$ Gauge - Higgs model this upper bound
cannot exceed $10 M_W$.}, and both triviality and Hierarchy scales approach
each other.

From the previous research we know that the phase diagram in the $\beta$ -
$\gamma$ plane of the lattice $SU(2)$ Gauge - Higgs for any fixed $\lambda$
resembles the phase diagram for the lattice Weinberg - Salam model. The only
difference is that in the $SU(2)$ Gauge - Higgs model the
confinement-deconfinement phase transition corresponding to the $U(1)$
constituents of the model is absent. The direct measurement of the renormalized
coupling $\beta_R$ shows \cite{1,2,3,4,5,6,7,8,9,10,11,12,13,14} that the line
of constant renormalized coupling constant (with the value close to the
experimental one) intersects the phase transition line. Also we know from the
direct measurements of $M_W$ in the $SU(2)$ Gauge - Higgs model that the
ultraviolet cutoff is increased when one is moving along this line from the
physical Higgs phase to the symmetric phase.

On the tree level the gauge boson mass in lattice units vanishes on the
transition surface at small enough $\lambda$. This means that the tree level
estimate predicts the appearance of an infinite ultraviolet cutoff at the
transition point for small $\lambda$. At infinite $\lambda$ the tree level
estimate gives nonzero values of lattice masses at the transition point. Our
numerical investigation of $SU(2)\otimes U(1)$ model (at $\lambda = 0.0025,
0.009, 0.001$) and previous calculations in the $SU(2)$ Gauge Higgs model (both
at finite $\lambda$ and at $\lambda = \infty$) showed that for the considered
lattice sizes renormalized masses do not vanish and the transition is either of
the first order or a crossover. (Actually, the situation, when the cutoff tends
to infinity at the position of the transition point means that there is a
second order phase transition.) The dependence on the lattice sizes for the
$SU(2)$ Gauge Higgs model was investigated, for example, in \cite{10}. Namely,
for $\beta = 8$, $\lambda \sim 0.00116$, where $M_H \sim M_W$, the correlation
lengths were evaluated at the transition points. For different lattice sizes
(from $12^3\times 28$ to $18^3 \times 36$) no change in correlation length was
observed \cite{10}.

In table 1 of \cite{BVZ2007} the data on the ultraviolet cutoff achieved in selected
lattice studies of the $SU(2)$ Gauge Higgs model are presented. Everywhere $\beta$ is
around $\beta \sim 8$ and the renormalized fine structure constant is around $\alpha \sim
1/110$. This table shows that the maximal value of the cutoff ${\Lambda} = \frac{\pi}{a}$
ever achieved in these studies is around $1.4$ Tev.

Thus the predictions on the value of $\Lambda_c$ given by our lattice study and
on the value $\Lambda_c^{0}$ given by the perturbation theory contradict with
each other. A possible explanation of this contradiction we suggested in
\cite{Z2009}. Namely, it was demonstrated that in the vicinity of the
transition  there exists the fluctuational region. Within this region the
application of the perturbation theory is limited. This situation is similar to
that of some phenomenological models that describe condensed matter
systems\footnote{One of the examples of such models is the Ginzburg - Landau
theory of superconductivity.}, where there exists the vicinity of the finite
temperature phase transition that is also called fluctuational region. In this
region the fluctuations of the order parameter become strong. The contribution
of these fluctuations to certain physical observables becomes larger than the
tree level estimate. Thus the perturbation theory in these models fails down
within the fluctuational region.

We find that there exists the vicinity of the phase transition between the
Higgs phase and the symmetric phase in the Weinberg - Salam model, where the
fluctuations of the scalar field become strong and the perturbation expansion
around trivial vacuum cannot be applied. According to the numerical results the
continuum theory is to be approached within the vicinity of the phase
transition, i.e. the cutoff is increased along the line of constant physics
when one approaches the point of the transition. That's why the conventional
prediction on the value of the cutoff admitted in the Standard Model based on
the perturbation theory may be incorrect.

In the present paper we proceed  the investigation \cite{Z2009} of  the model
at the value of the scalar self coupling $\lambda = 0.009$ (corresponds to the
Higgs boson mass around $270$ Gev in the vicinity of the phase transition),
bare Weinberg angle $\theta_W = 30^o$, and bare fine structure constant around
$1/150$. The results presented now correspond to essentially larger lattices
than that of used in \cite{Z2009}. Namely, in \cite{Z2009} main results
correspond to lattices $8^3\times 16$; some results were checked on the lattice
$12^3\times 16$; two points were checked on the lattice $16^4$. Now our main
results are obtained on the lattice $16^4$ while the results at the transition
point were checked on the lattice $20^3\times 24$.

In addition we investigate the model at the value of the scalar self coupling
$\lambda = 0.0025$,  bare Weinberg angle $\theta_W = 30^o$, and bare fine
structure constant around $\alpha_0 \sim 1/150$. These values of couplings
correspond to the Higgs boson mass around $150$ Gev in the vicinity of the
phase transition. The results are obtained using lattices $8^3\times 16$,
$12^3\times 16$, and $16^4$. We also present results for $\lambda = 0.001$,
$\theta_W = 30^o$,  $\alpha_0 \sim 1/150$. These values of couplings correspond
to the Higgs boson mass around $100$ Gev. The results are obtained using
lattices $8^3\times 16$, $12^3\times 16$.

 It is worth mentioning that far from
the transition point the renormalized fine structure constant slowly approaches
the tree level estimate. Contrary to the maximal value of the cutoff the
renormalized fine structure constant depends on the lattice size. And for the
larger lattice the value of $\alpha_R$ is closer to the tree level estimate
than for the smaller one. For example, for $\beta = 12, \gamma \sim 1, \lambda
= 0.001$ (far from the transition point) on the lattice $8^3\times 16$ the
value of $\alpha_R$ is around $1/130$ while on the lattice $12^3\times 16$ it
is around $1/140$. Within the fluctuational region the deviation from tree
level estimate becomes essentially strong. For example, for $\lambda = 0.009,
\gamma = 0.274$ (near the transition point) the renormalized value of
$\alpha_R$ calculated on the lattice $8^3\times 16$ is around $1/99$ while on
the lattice $20^3\times 24$ its value is around $1/106$. As it is seen from our
numerical results and as it will be explained in the Conclusions we guess the
mentioned finite volume effects present in  the value of renormalized $\alpha$
do not affect the main observables we considered like the value of $\Lambda_c$
and the Nambu monopole density.

We calculate the constraint effective potential $V(|\Phi|)$ for the Higgs field
$\Phi$. In the physical Higgs phase this potential has a minimum at a certain
nonzero value $\phi_m$ of $|\Phi|$. This shows that the spontaneous breakdown
of the Electroweak symmetry takes place as it should. However, there exists the
vicinity of the phase transition, where the fluctuations of the Higgs field are
of the order of $\phi_m$ while the hight of the "potential
barrier"\footnote{The meaning of the words "potential barrier" here is
different from that of the one - dimensional quantum mechanics as here
different minima of the potential form the three - dimensional sphere while in
usual $1D$ quantum mechanics with the similar potential there are two separated
minima with the potential barrier between them. Nevertheless we feel it
appropriate to use the chosen terminology as the value of the "potential
barrier hight" measures the difference between the potentials with and without
spontaneous symmetry breaking. }  $H = V(0) - V(\phi_m)$ is of the order of
$V(\phi_m + \delta \phi)-V(\phi_m)$, where $\delta \phi$ is the fluctuation of
$|\Phi|$. We expect that in this region the perturbation expansion around
trivial vacuum $\Phi = (\phi_m,0)^T$ cannot be applied. This region of the
phase diagram is called the fluctuational region (FR).

The nature of the fluctuational region is illustrated by the behavior of quantum Nambu
monopoles \cite{Nambu,Chernodub_Nambu}. We show that their lattice density increases when
the phase transition point is approached. Within the FR these objects are so dense that
it is not possible at all to speak of them as of single monopoles \footnote{It has been
shown in \cite{VZ2008} that at the infinite value of the scalar self coupling $\lambda =
\infty$ moving along the line of constant physics we reach the point on the phase diagram
where the monopole worldlines begin to percolate. This point was found to coincide
roughly with the position of the transition between the physical Higgs phase and the
unphysical symmetric phase of the lattice model. This transition is a crossover and the
ultraviolet cutoff achieves its maximal value around $1.4$ Tev at the transition
point.}. Namely, within this region the average distance between the Nambu monopoles is
of the order of their size. Such complicated configurations obviously have nothing to do
with the conventional vacuum used in the continuum perturbation theory.

\section{The lattice model under investigation}

The lattice Weinberg - Salam Model without fermions contains  gauge field ${\cal U} = (U,
\theta)$ (where $ \quad U
 \in SU(2), \quad e^{i\theta} \in U(1)$ are
realized as link variables), and the scalar doublet $ \Phi_{\alpha}, \;(\alpha = 1,2)$
defined on sites.

The  action is taken in the form
\begin{eqnarray}
 S & = & \beta \!\! \sum_{\rm plaquettes}\!\!
 ((1-\mbox{${\small \frac{1}{2}}$} \, {\rm Tr}\, U_p )
 + \frac{1}{{\rm tg}^2 \theta_W} (1-\cos \theta_p))+\nonumber\\
 && - \gamma \sum_{xy} Re(\Phi^+U_{xy} e^{i\theta_{xy}}\Phi) + \sum_x (|\Phi_x|^2 +
 \lambda(|\Phi_x|^2-1)^2), \label{S}
\end{eqnarray}
where the plaquette variables are defined as $U_p = U_{xy} U_{yz} U_{wz}^* U_{xw}^*$, and
$\theta_p = \theta_{xy} + \theta_{yz} - \theta_{wz} - \theta_{xw}$ for the plaquette
composed of the vertices $x,y,z,w$. Here $\lambda$ is the scalar self coupling, and
$\gamma = 2\kappa$, where $\kappa$ corresponds to the constant used in the investigations
of the $SU(2)$ gauge Higgs model. $\theta_W$ is the Weinberg angle.

Bare fine structure
 constant $\alpha$ is expressed through $\beta$ and $\theta_W$ as $\alpha = \frac{{\rm tg}^2 \theta_W}{\pi \beta(1+{\rm tg}^2
\theta_W)}$. In order to demonstrate this we consider naive continuum limit of (\ref{S}).
We set
\begin{eqnarray}
  \quad U_{x,\mu} = e^{iA_{\mu}(x)a}, \quad e^{i\theta_{x,\mu}} = e^{iB_{\mu}(x)a}
\end{eqnarray}
Here $a$ is the lattice spacing. The field $B_{\mu}=\frac{\tilde{B_{\mu}}}{2}$, where
$\tilde{B_{\mu}}$ - is the conventional $U(1)$ field while $A_{\mu}$ is the conventional
$SU(2)$ field. In continuum limit (\ref{S}) must become
\begin{eqnarray}
 S_g & = & \int d^4x
 \{\frac{1}{2g_2^2}  {\rm Tr}\, [ 2 \times  \sum_{i>j}G^2_{ij}]
 + \frac{1}{4g_1^2}  [ 2 \times \sum_{i>j}\tilde{F}^2_{ij}]
  \},\label{Act0c}
\end{eqnarray}
Here $\tilde{F}_{ij} = \partial_{i}\tilde{B}_j - \partial_{j}\tilde{B}_i = 2
(\partial_{i}{B}_j - \partial_{j}{B}_i) = 2 F_{ij}$, ${G}_{ij} =
\partial_{i}{A}_j -
\partial_{j}{A}_i - i[A_i,A_j]$.
We also have the following correspondence between the plaquette variables and the field
strengths:
\begin{eqnarray}
   \quad {\rm Tr} U_{x,\mu\nu} &=& {\rm Tr}[1-\frac{1}{2}G^2_{\mu
\nu}a^4],
 \nonumber\\  \quad {\rm cos} \, N {\theta_{x,\mu\nu}} &=& [1-\frac{N^2}{2}{F}^2_{\mu \nu}a^4]
\end{eqnarray}

Now in order to clarify the correspondence between constants $g_{1,2}$ and $\beta$ we
must substitute the expressions for the field strengths to (\ref{S}) and compare it to
(\ref{Act0c}). We have:
\begin{eqnarray}
 \frac{1}{g^2_1} = \frac{1}{4{\rm tg}^2 \theta_W} \times  \beta , \quad \frac{1}{g^2_2} =
 \beta/4
\end{eqnarray}

Thus
\begin{eqnarray}
 {\rm tg} \theta_W &=& \frac{g_1}{g_2} ,\nonumber\\
   \quad \alpha &=&
 \frac{e^2}{4\pi}= \frac{[\frac{1}{g^2_1}+\frac{1}{g^2_2}]^{-1}}{4\pi}= \frac{{\rm tg}^2 \theta_W}{\pi \beta(1+{\rm tg}^2
\theta_W)}
\end{eqnarray}

We consider the region of the phase diagram with  $\beta \sim 12$ and $\theta_W \sim
\pi/6$. Therefore, bare couplings are ${\rm sin}^2 \theta_W \sim 0.25$; $\alpha \sim
\frac{1}{150}$. These values are to be compared with the experimental ones ${\rm sin}^2
\theta_W(100 {\rm Gev}) \sim 0.23$; $\alpha(100 {\rm Gev}) \sim \frac{1}{128}$.

The simulations were performed on lattices of sizes $8^3\times 16$, $12^3\times
16$. For $\lambda = 0.0025, 0.009$ we investigate the system on the lattice
$16^4$. The transition point at $\lambda = 0.009$ was checked using the larger
lattice ($20^3\times 24$). In order to simulate the system we used Metropolis
algorithm. The acceptance rate is kept around $0.5$ via the automatical self -
tuning of the suggested distribution of the fields. At each step of the
suggestion the random value is added to the old value of the scalar field while
the old value of Gauge field is multiplied by random $SU(2)\otimes U(1)$
matrix. We use Gaussian distribution both for the random value added to the
scalar field and the parameters of the random matrix multiplied by the lattice
Gauge field. We use two independent parameters for these distributions: one for
the Gauge fields and another for the scalar field. The program code has been
tested for the case of frozen scalar field. And the results of the papers
\cite{VZ2008} are reproduced. We also have tested our code for the $U(1)$ field
frozen and repeat the results of \cite{Montvayold}.  Far from the transition
point  the autocorrelation time for the gauge fields is estimated as about
$N^g_{auto} \sim 500$ Metropolis steps. In the vicinity of the transition point
the autocorrelation
  time is several times larger and is about $N^g_{auto} \sim 1500$ Metropolis steps. (The correlation between the values of the gauge field is less
than $3 \%$ for the configurations separated by $N^g_{auto}$ Metropolis steps.
Each metropolis step consists of the renewing the fields over all the lattice.)
The autocorrelation time for the scalar field is essentially smaller than for
the gauge fields and is of the order of $N^{\phi}_{auto} \sim 20$. The
estimated time for preparing the equilibrium starting from the cold start far
from the phase transition within the Higgs phase is about $18000$ Metropolis
steps for the considered values of couplings. At the same time near the phase
transition and within the symmetric phase the estimated time for preparing the
equilibrium is up to $3$ times larger.

\section{The tree level estimates of lattice quantities}

At finite $\lambda$ the line of constant renormalized $\alpha$ is not a line of constant
physics, because the mass of the Higgs boson depends on the position on this line. Thus,
in order to investigate the line of constant physics  one should vary $\lambda$ together
with $\gamma$ to keep the ratio of lattice masses $M_H/M_W$ constant.

In order to obtain the tree level estimates let us rewrite the lattice action in an
appropriate way. Namely, we define the scalar field $\tilde{\Phi} =
\sqrt{\frac{\gamma}{2}} \Phi$. We have:

\begin{eqnarray}
 S & = & \beta \!\! \sum_{\rm plaquettes}\!\!
 ((1-\mbox{${\small \frac{1}{2}}$} \, {\rm Tr}\, U_p )
 + \frac{1}{{\rm tg}^2 \theta_W} (1-\cos \theta_p))+\nonumber\\
 && + \sum_{xy} |\tilde{\Phi}_x - U_{xy} e^{i\theta_{xy}}\tilde{\Phi}_y|^2 + \sum_x (\mu^2 |\tilde{\Phi}_x|^2 +
 \tilde{\lambda} |\tilde{\Phi}_x|^4) + \omega , \label{S2}
\end{eqnarray}
where $\mu^2 = - 2(4+(2\lambda-1)/\gamma)$, $\tilde{\lambda} =
4\frac{\lambda}{\gamma^2}$, and $\omega = \lambda V$. Here  $V = L^4$ is the lattice
volume, and $L$ is the lattice size.

For negative $\mu^2$ we fix Unitary gauge $\tilde{\Phi}_2=0$, ${\rm Im}\, \tilde{\Phi}_1
= 0$, and introduce the vacuum value of $\tilde{\Phi}$: $v =
\frac{|\mu|}{\sqrt{2\tilde{\lambda}}}$. We also introduce the scalar field $\sigma$
instead of $\tilde{\Phi}$: $\tilde{\Phi}_1 = v + \sigma$. We denote $V_{xy} =
(U^{11}_{xy}e^{i\theta_{xy}} - 1)$, and obtain:
\begin{eqnarray}
 S & = & \beta \!\! \sum_{\rm plaquettes}\!\!
 ((1-\mbox{${\small \frac{1}{2}}$} \, {\rm Tr}\, U_p )
 + \frac{1}{{\rm tg}^2 \theta_W} (1-\cos \theta_p))+\nonumber\\
 && + \sum_{xy} ((\sigma_x - \sigma_y)^2 + |V_{xy}|^2 v^2)  + \sum_x 2|\mu|^2 \sigma_x^2 \nonumber\\
 && + \sum_{xy} ((\sigma^2_y+2v \sigma_y)|V_{xy}|^2 - 2(\sigma_x - \sigma_y){\rm Re} V_{xy} (\sigma_y +v) ) + \nonumber\\
 && + \sum_x  \tilde{\lambda} \sigma_x^2 (\sigma_x^2 + 4 v \sigma_x) +  \tilde{\omega} , \label{S2}
\end{eqnarray}
where $\tilde{\omega} = \omega - \tilde{\lambda} v^4 V$.

Now we easily derive the tree level estimates:
\begin{eqnarray}
M_H &=& \sqrt{2}|\mu| = 2\sqrt{4+(2\lambda-1)/\gamma}; \nonumber\\
M_W &=& \sqrt{2} \frac{v}{\sqrt{\beta}} =  \sqrt{\frac{\gamma(4\gamma+2\lambda-1)}{2\lambda\beta}}; \nonumber\\
M_W &=& {\rm cos}\theta_W M_Z\nonumber\\
M_H/M_W &=& \sqrt{8\lambda \beta/\gamma^2};\nonumber\\
\Lambda &=&\pi \sqrt{\frac{2\lambda\beta}{\gamma(4\gamma+2\lambda-1)}} \, [80\, {\rm
GeV}];\label{tree}
\end{eqnarray}
The fine structure constant is given by $\alpha = \frac{{\rm tg}^2 \theta_W}{\pi
\beta(1+{\rm tg}^2 \theta_W)}$ and does not depend on $\lambda$ and $\gamma$. From
(\ref{tree}) we learn that at the tree level LCP on the phase diagram corresponds to
fixed $\beta = \frac{{\rm tg}^2 \theta_W}{\pi \alpha(1+{\rm tg}^2 \theta_W)} \sim 10 $
and $\eta = M_H/M_W$, and is given by the equation $\lambda(\gamma) =
\frac{\eta^2}{8\beta} \gamma^2$.

The important case is $\lambda = \infty$, where the tree level estimates give
\begin{eqnarray}
M_H &=& \infty; \nonumber\\
M_W &=& \sqrt{\frac{\gamma}{\beta}}; \nonumber\\
M_Z &=& \sqrt{\frac{\gamma}{\beta}}{\rm cos}^{-1}\theta_W; \nonumber\\
\Lambda &=& \pi \sqrt{\frac{\beta}{\gamma}} \, [80\, {\rm GeV}];\label{treei}
\end{eqnarray}

In the $SU(2)$ gauge Higgs model for the small values of $\lambda << 0.1$ the tree level
estimate for $M_H/M_W$ gives values that differ from the renormalized ratio by about
20\%\cite{11}.
The tree level estimate for the ultraviolet cutoff is about $1$ TeV at $\lambda =
\infty,\gamma = 1, \beta = 15$ that is not far from the numerical result given in
\cite{VZ2008}. In the $SU(2)$ Gauge Higgs model at $\lambda = \infty$ the critical
$\gamma_c = 0.63$ for $\beta = 8$ \cite{14}. At
 this point the tree level estimate gives $\Lambda = 0.9$ Tev while the direct measurements
 give $\Lambda \in [0.8; 1.5]$ Tev for values of $\gamma \in [0.64; 0.95]$ \cite{14}. The investigations of
the $SU(2)$ Gauge Higgs model showed that a consideration of finite $\lambda$ does not
change much the estimate for the gauge boson mass. However, at finite $\lambda$ and
values of $\gamma$ close to the phase-transition point the tree level formula does not
work at all.

The tree level estimate for the critical $\gamma$ is $\gamma_c = (1-2\lambda)/4$. At
small $\lambda$ this formula gives values that are close to the ones obtained by the
numerical simulations \cite{12,13,14}. In particular, $\gamma_c \rightarrow 0.25$
($\kappa_c \rightarrow 0.125$) at $\lambda << 1$. However, this formula clearly does not
work for $\lambda > 1/2$. From \cite{Montvay,12,13,14} we know that the critical coupling
in the $SU(2)$ Gauge Higgs model is about $2 - 4$ times smaller for $\lambda =0$ than for
$\lambda = \infty$.

Tree level estimate predicts that there is the second order phase transition.
This means that according to the tree level estimate the value of the cutoff at
the transition point is infinite. Our numerical simulations, however, show that
the cutoff remains finite and the transition is, most likely, a crossover at
the considered values of $\theta_W$, $\lambda$ and $\beta$.

\section{Nambu monopoles}

In this section we remind the reader what is called Nambu monopole \cite{Nambu}. First
let us define the continuum Electroweak fields as they appear in the Weinberg-Salam
model. The continuum scalar doublet is denoted as $\Phi$. The $Z$-boson field $Z^{\mu}$
and electromagnetic field $A_{\rm EM}^{\mu}$ are defined as
\begin{eqnarray}
 Z^{\mu} = - \frac{1}{\sqrt{\Phi^+ \Phi}} \Phi^+ A^{\mu} \Phi -  B^{\mu},
\nonumber\\
 A_{\rm EM}^{\mu} =  2 B^{\mu}  + 2 \,{\rm sin}^2\, \theta_W
 Z^{\mu},\label{FSM}
\end{eqnarray}
where $A^{\mu}$ and $B^{\mu}$ are the corresponding $SU(2)$ and $U(1)$ gauge fields of
the Standard Model.

After fixing the unitary gauge $\Phi_2=const.$, $\Phi_1 = 0$ we have
\begin{eqnarray}
 Z^{\mu} =  \frac{g_z}{2}[\frac{\tilde{A_3}^{\mu}}{g_2}{\rm cos}\theta_W -  \frac{\tilde{B}^{\mu}}{g_1}{\rm sin}\theta_W] = \frac{1}{2}\tilde{Z}^{\mu},
\nonumber\\
 A_{\rm EM}^{\mu} =  e[\frac{\tilde{A_3}^{\mu}}{g_2}{\rm sin}\theta_W +  \frac{\tilde{B}^{\mu}}{g_1}{\rm cos}\theta_W] = \tilde{A}^{\mu},
\end{eqnarray}
where $\frac{\tilde{A}_3}{g_2} = \frac{1}{g_2}{\rm Tr}\, A \sigma^3$,
$\frac{\tilde{B}}{g_1} = 2 B/g_1$,  $\frac{\tilde{Z}}{g_z}$, $\frac{\tilde{A}}{e}$ -
conventional Standard Model fields, and $g_z = \sqrt{g_1^2+g_2^2}$.

Nambu monopoles are defined as the endpoints of the $Z$-string \cite{Nambu}. The
$Z$-string is the classical field configuration that represents the object, which is
characterized by the magnetic flux extracted from the $Z$-boson field. Namely, for a
small contour $\cal C$ winding around the $Z$ - string one should have
\begin{equation}
 \int_{\cal C} Z^{\mu} dx^{\mu} \sim 2\pi;\,
 \int_{\cal C} A_{\rm EM}^{\mu} dx^{\mu} \sim 0;\,
 \int_{\cal C} B^{\mu} dx^{\mu} \sim 2\pi {\rm sin}^2\, \theta_W .
\end{equation}
The string terminates at the position of the Nambu monopole. The hypercharge flux is
supposed to be conserved at that point. Therefore, a Nambu monopole carries
electromagnetic flux $4\pi {\rm sin}^2\, \theta_W$. The size of Nambu monopoles was
estimated \cite{Nambu} to be of the order of the inverse Higgs mass, while its mass
should be of the order of a few TeV. According to \cite{Nambu} Nambu monopoles may appear
only in the form of a bound state of a monopole-antimonopole pair.

In lattice theory the following variables are considered as creating the $Z$ boson:
\begin{equation} Z_{xy} = Z^{\mu}_{x} \;
 = - {\rm sin} \,[{\rm Arg} (\Phi_x^+U_{xy} e^{i\theta_{xy}}\Phi_y) ]. \label{Z1}
\end{equation}
and:
\begin{equation} Z^{\prime}_{xy} = Z^{\mu}_{x} \;
 = - \,[{\rm Arg} (\Phi_x^+U_{xy} e^{i\theta_{xy}}\Phi_y) ]. \label{Z1_}
\end{equation}

The classical solution corresponding to a $Z$-string should be formed around
the $2$-dimensional topological defect which is represented by the
integer-valued field defined on the dual lattice $ \Sigma = \frac{1}{2\pi}^*([d
Z^{\prime}]_{{\rm mod} 2\pi} - d Z^{\prime})$. (Here we used the notations of
differential forms on the lattice. For a definition of those notations see, for
example, ~\cite{forms}. Lattice field $Z^\prime$ is defined in Eg.
(\ref{Z1_}).) Therefore, $\Sigma$ can be treated as the worldsheet of a {\it
quantum} $Z$-string \cite{Chernodub_Nambu}. Then, the worldlines of quantum
Nambu monopoles appear as the boundary of the $Z$-string worldsheet: $ j_Z =
\delta \Sigma $.

For historical reasons in lattice simulations we fix unitary gauge $\Phi_2 = 0$; $\Phi_1
\in {\cal R}$; $\Phi_1 \ge 0$ (instead of the usual $\Phi_1 = 0$; $\Phi_2 \in {\cal R}$),
and the lattice Electroweak theory becomes a lattice $U(1)$ gauge theory with the $U(1)$
gauge field
\begin{equation}
 A_{xy}  =  A^{\mu}_{x} \;
 = \,[Z^{\prime}  + 2\theta_{xy}]  \,{\rm mod}
 \,2\pi, \label{A}
\end{equation}
(The usual lattice Electromagnetic field is related to $A$ as $ A_{\rm EM}  = A -
Z^{\prime} + 2 \,{\rm sin}^2\, \theta_W Z^{\prime}$.) One may try to extract monopole
trajectories directly from $A$. The monopole current is given by
\begin{equation}
 j_{A} = \frac{1}{2\pi} {}^*d([d A]{\rm mod}2\pi)
\label{Am}
\end{equation}
Both $j_Z$, and $j_A$ carry magnetic charges. That's why it is important to find the
correspondence between them.

In continuum notations we have
\begin{equation}
 A^{\mu}  =  Z^{\mu} + 2 B^{\mu},
\end{equation}
where $B$ is the hypercharge field. Its strength is divergenceless. As a result in
continuum theory the net $Z$ flux emanating from the center
 of the monopole is equal to the net $A$ flux.
(Both $A$ and $Z$ are undefined inside the monopole.)  This means that in the continuum
limit the position of the Nambu monopole must coincide
 with the position of the antimonopole extracted from the field $A$.
Therefore, one can consider Eq.~(\ref{Am}) as another definition of a quantum Nambu
monopole \cite{VZ2008}. Actually, in our numerical simulations we use the definition of
Eq. (\ref{Am}).

\section{Phase diagram}

\begin{figure}
\begin{center}
 \epsfig{figure=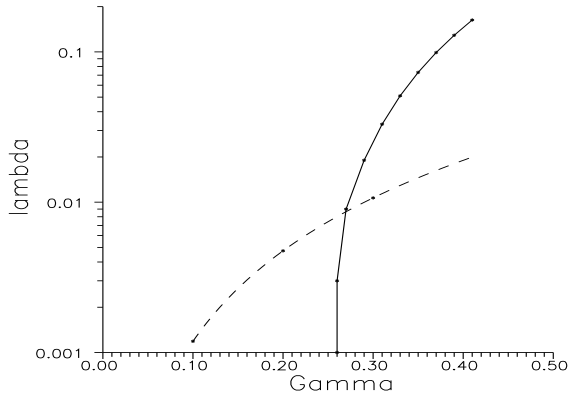,height=60mm,width=80mm,angle=0}
\caption{\label{fig.2} The phase diagram of the model in the
 $(\gamma, \lambda)$-plane at $\beta = 12$. The dashed line is the tree - level
  estimate for the line of constant physics correspondent to bare $M^0_H = 270$ Gev. The continuous line
  is the line of phase transition between the physical Higgs phase and the unphysical symmetric phase (statistical errors for
  the values of $\gamma$ at each $\lambda$ on this line are about $0.005$).   }
\end{center}
\end{figure}

In our lattice study we fix bare $\theta_W = \pi/6$. Then in the three -
dimensional ($\beta, \gamma, \lambda$) phase diagram the transition surfaces
are two - dimensional. The lines of constant physics on the tree level are the
lines ($\frac{\lambda}{\gamma^2} = \frac{1}{8 \beta} \frac{M^2_H}{M^2_W} = {\rm
const}$; $\beta = \frac{1}{4\pi \alpha}={\rm const}$). We suppose that in the
vicinity of the transition  the deviation of the lines of constant physics from
the tree level estimate may be significant. However,  qualitatively their
behavior is the same. Namely, the cutoff is increased along the line of
constant physics when $\gamma$ is decreased and the maximal value of the cutoff
is achieved at the transition point. Nambu monopole density in lattice units is
also increased when the ultraviolet cutoff is increased.

\begin{figure}
\begin{picture}(0,0)(0,0)
\put(-120,-20){$\chi$} \put(80,-190){$\Large \gamma$}
\end{picture}
\begin{center}
 \epsfig{figure=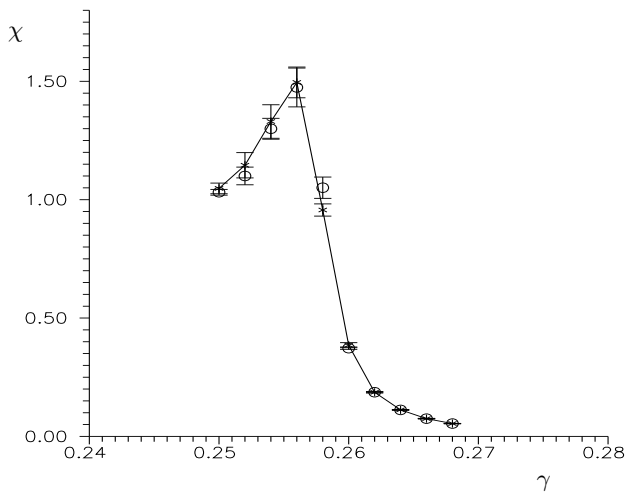,height=60mm,width=80mm,angle=0}
 \end{center}
 \begin{center}
 \caption{\label{fig.6__} Susceptibility $\chi = \langle H^2 \rangle -
\langle H\rangle^2$ (for $H_x = \sum_{y} Z^2_{xy}$) as a function of $\gamma$
at $\lambda =0.001$ and $\beta = 12$. Circles correspond to the lattice
$12^3\times16$. Crosses correspond to the lattice $8^3\times 16$. }
\end{center}
\end{figure}

At $\beta = 12$ (corresponds to bare $\alpha \sim 1/150$) the phase diagram is
represented on Fig. \ref{fig.2}. This diagram is obtained, mainly, using the
lattice $8^3\times 16$. Some regions ($\lambda = 0.009,0.0025, 0.001$),
however, were checked using larger lattices. According to our data there is no
dependence of the diagram on the lattice size. The physical Higgs phase is
situated right to the transition line. The position of the transition
$\gamma_c(\lambda)$ is localized here at the point where the susceptibility
extracted from the Higgs field creation operator
 achieves its maximum.
 We use the
susceptibility
\begin{equation}
\chi = \langle H^2 \rangle - \langle H\rangle^2 \label{chiH}
\end{equation}
 extracted from $H = \sum_{y}
Z^2_{xy}$ (see, for example,  Fig. \ref{fig.6__}). We observe no difference
between the values of the susceptibility calculated using the lattices of
different sizes. This indicates that the  transition at $\gamma_c$ is a
crossover. Indeed we find that gauge boson masses do not vanish in a certain
vicinity of $\gamma_c$ even within the symmetric phase. In the next section we
shall see that within the statistical errors $\gamma_c$ coincides with the
value of $\gamma$, where the scalar field condensate disappears.  Actually,
there also exist two other crucial points: $\gamma_{c0}(\lambda) <
\gamma_c(\lambda) < \gamma_{c2}(\lambda)$ (say, at $\lambda = 0.001$ we have
$\gamma_{c0} = 0.252\pm 0.001$, $\gamma_{c} = 0.256\pm 0.001$, $\gamma_{c2} =
0.258\pm 0.001$, see the next sections for the details). $\gamma_{c2}$ denotes
the boundary of the fluctuational region.  At $\gamma_{c0}$ the extrapolation
of the dependence of lattice $Z$ - boson mass $M_Z(\gamma)$ on $\gamma$
indicates that $M_Z(\gamma_{c0})$ may vanish. In the symmetric phase the
perturbation theory predicts vanishing of the gauge boson masses. Therefore,
supposition that $M_Z$ vanishes at a certain point is very natural. The
perturbation theory also predicts that the mass parameter present in the
effective action for the scalar field vanishes at the point of the transition
between Higgs phase and the symmetric phase. Our analysis shows that at the
point, where the scalar field condensate disappears lattice $M_H$ does not
vanish. However, it may vanish, in principle, at some other point. If both
$M_Z$ and $M_H$ vanish simultaneously at $\gamma_{c0}$, at this point the model
becomes scale invariant and formal continuum limit of the lattice model can be
achieved at $\gamma_{c0}$. This point may then appear as the point of the
second order phase transition. Near $\gamma_{c0}$ the fluctuations of the gauge
boson correlator are strong and at the present moment we do not make definite
conclusions on the behavior of the system at $\gamma_{c0}$. However, the
calculated susceptibilities do not have peaks at this point that is an indirect
indication that the real second order phase transition cannot appear at
$\gamma_{c0}$. It is worth mentioning that within the region $(\gamma_{c0},
\gamma_c)$ the scalar field is not condensed. That's why we guess this region
has nothing to do with real continuum physics.

We investigated carefully the region $\gamma \ge \gamma_c$ for $\lambda =
0.001, 0.0025, 0.009$. We observe that for $\gamma_c < \gamma < \gamma_{c2}$
Nambu monopoles dominate vacuum and the usual perturbation theory cannot be
applied. For this reason, most likely, the interval $(\gamma_c, \gamma_{c2})$
also has no connection with the conventional continuum Electroweak theory. At
the same time for $\gamma
>> \gamma_{c2}$ the behavior of the system is close to what one would expect basing on the
usual perturbative continuum Weinberg - Salam model. It is worth mentioning
that the value of the renormalized Higgs boson mass does not deviate
significantly from its bare value near the transition point $\gamma_c$. For
example, for $\lambda$ around $0.009$ and $\gamma = 0.274$ bare value of the
Higgs mass is around $270$ Gev while the observed renormalized value is  $300
\pm 70$ Gev.

\section{Effective constraint potential}

\begin{figure}
\begin{center}
 \epsfig{figure=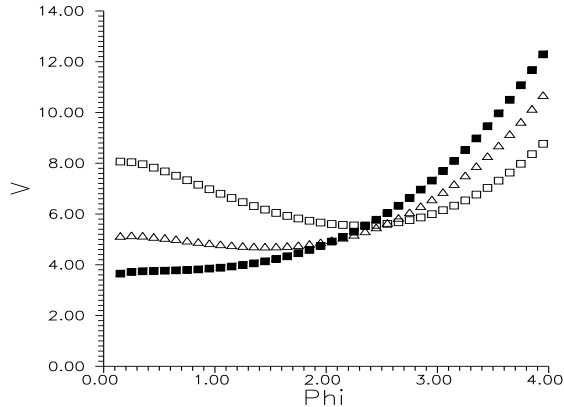,height=60mm,width=80mm,angle=0}
\caption{\label{fig.1} The effective constraint potential at $\lambda =0.009$ and $\beta
= 12$. Black squares correspond to $\gamma_c = 0.273$. Empty squares correspond to
$\gamma =0.29$. Triangles correspond to $\gamma = 0.279$. The error bars are about of the
same size as the symbols used. }
\end{center}
\end{figure}

\begin{figure}
\begin{center}
 \epsfig{figure=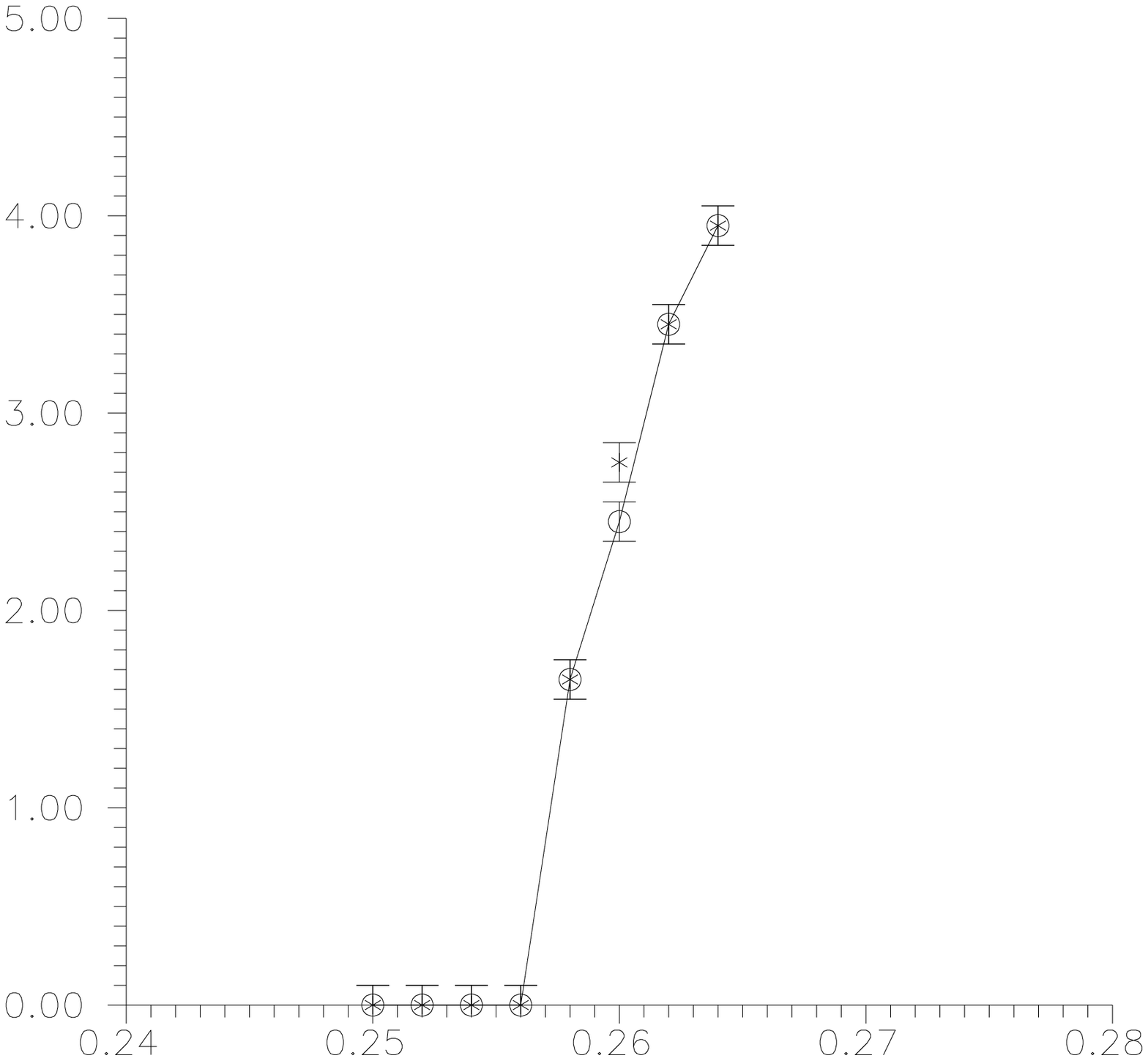,height=60mm,width=80mm,angle=0}
\end{center}
\begin{picture}(0,0)(0,0)
\put(-125,170){$\phi_m$} \put(80,10){$\Large \gamma$}
\end{picture}
\begin{center}
\caption{\label{fig.4_} $\phi_m$  as a function of $\gamma$ at $\lambda =0.001$
and $\beta = 12$. Circles correspond to lattice $8^3\times 16$. Crosses
correspond to lattice $12^3\times 16$. }
\end{center}
\end{figure}

We have calculated the constraint effective potential for $|\Phi|$ using the histogram
method. The calculations have been performed on the lattice $8^3\times 16$. The
probability $h(\phi)$ to find the value of $|\Phi|$ within the interval
$[\phi-0.05;\phi+0.05)$ has been calculated for $\phi = 0.05 + N*0.1$, $N = 0,1,2, ...$
This probability is related to the effective potential as $ h(\phi) = \phi^3
e^{-V(\phi)}$. That's why we extract the potential from $h(\phi)$ as
\begin{equation}
V(\phi) = - {\rm log}\, h(\phi) + 3 \, {\rm log} \, \phi \label{CEP}
\end{equation}
(See Fig. \ref{fig.1}.) It is worth mentioning that $h(0.05)$ is calculated as
the probability to find the value of $|\Phi|$ within the interval $[0;0.1]$.
Within this interval ${\rm log}\, \phi$ is ill defined. That's why we exclude
the point $\phi = 0.05$ from our data. Instead we calculate $V(0)$ using the
extrapolation of the data at $0.15 \le \phi \le 2.0$. The extrapolation is
performed using the polynomial fit with the powers of $\phi$ up to the third
(average deviation of the fit from the data is around $1$ per cent). Next, we
introduce the useful quantity $H = V(0) - V(\phi_m)$, which is called the
potential barrier hight (here $\phi_m$ is the point, where $V$ achieves its
minimum).

\begin{figure}
\begin{center}
 \epsfig{figure=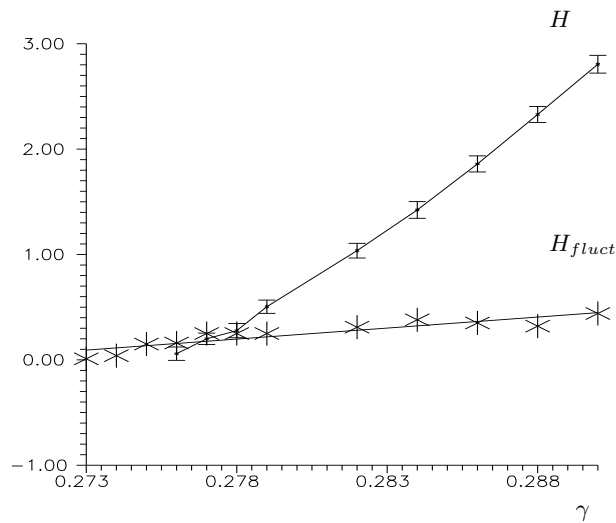,height=60mm,width=80mm,angle=0}
\end{center}
\begin{picture}(0,0)(0,0)
\put(90,195){$H$} \put(90,110){$H_{fluct}$} \put(100,10){$\Large \gamma$}
\end{picture}
\begin{center}
\caption{\label{fig.3_} $H$ (points) vs. $H_{fluct}$ (stars)  as a function of
$\gamma$ at $\lambda =0.009$ and $\beta = 12$. Statistical errors for
$H_{fluct}$ are about of the same size as the symbols used.  }
\end{center}
\end{figure}

\begin{figure}
\begin{center}
 \epsfig{figure=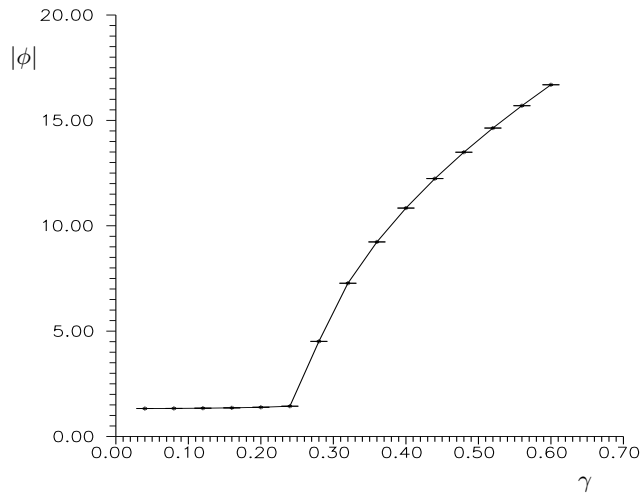,height=60mm,width=80mm,angle=0}
\end{center}
\begin{picture}(0,0)(0,0)
\put(-125,170){$|\phi|$} \put(90,10){$\Large \gamma$}
\end{picture}
\begin{center}
\caption{\label{fig.2_} Mean value of $|\phi|$ as a function of $\gamma$ at
$\lambda =0.0025$ and $\beta = 12$. (Lattice $8^3\times 16$.) }
\end{center}
\end{figure}

As an example we represent on  Fig. \ref{fig.4_} the values of $\phi_m$ for
$\lambda = 0.001$, $\beta = 12$.  On Fig. \ref{fig.3_} we represent the values
of $H$ for $\lambda = 0.009$, $\beta = 12$. One can see that the values of
$\phi_m$ and $H$ increase when $\gamma$ is increased.  The maximum of the
susceptibility constructed of the Higgs field creation operator $H_x = \sum_{y}
Z^2_{xy}$ (see, for example,  Fig. \ref{fig.6__}) coincides with the point,
where $\phi_m$ vanishes within the statistical errors.
 We localize the
position of the transition points at the points where $\phi_m$ vanishes:
$\gamma_c = 0.274\pm 0.001$ at $\lambda = 0.009$;  $\gamma_c = 0.26 \pm 0.001$
at $\lambda = 0.0025$; and $\gamma_c = 0.256 \pm 0.001$ at $\lambda = 0.001$.

\begin{figure}
\begin{center}
 \epsfig{figure=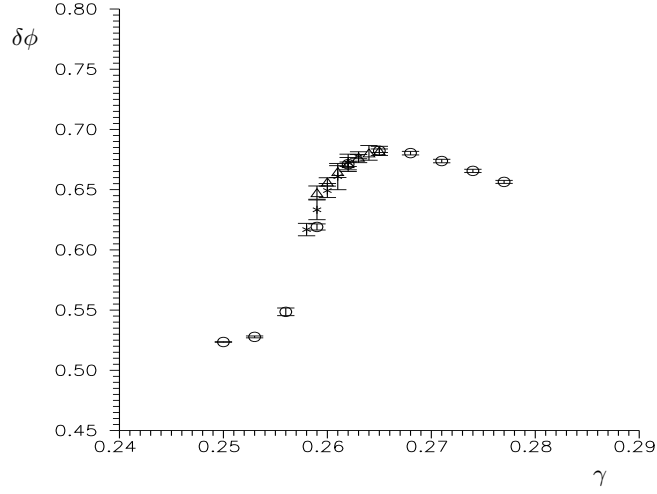,height=60mm,width=80mm,angle=0}
\end{center}
\begin{picture}(0,0)(0,0)
\put(-130,175){$\delta \phi$} \put(90,10){$\Large \gamma$}
\end{picture}
\begin{center}
\caption{\label{fig.6_2_3} Fluctuation $\delta \phi$ as a function of $\gamma$
at $\lambda =0.0025$ and $\beta = 12$. Circles correspond to the lattice
$8^3\times16$. Crosses correspond to the lattice $12^3\times 16$. Triangles
correspond to the lattice $16^4$. Transition point is $\gamma_c = 0.261\pm
0.001$; it is clear that the maximum of $\delta \phi$ is shifted to larger
values of $\gamma$. }
\end{center}
\end{figure}

The maximum of the scalar field fluctuation (see, for example,  Fig.
\ref{fig.6_2_3}) is shifted  to larger values of $\gamma$ than the transition
point. Again we do not observe any difference in $\delta \phi$ for the
considered lattice sizes. This also indicates that the transition at these
values of $\lambda$ is a crossover.

It is important to understand which value of barrier hight can be considered as
small and which value can be considered as large. Our suggestion is to compare
$H = V(0) - V(\phi_m)$ with $H_{\rm fluct} = V(\phi_m + \delta \phi) -
V(\phi_m)$, where $\delta \phi$ is the fluctuation of $|\Phi|$. From Fig.
\ref{fig.3_} it is clear that there exists the value of $\gamma$ (we denote it
$\gamma_{c2}$) such that at $\gamma_c < \gamma < \gamma_{c2}$ the barrier hight
$H$ is of the order of $H_{\rm fluct}$ while for $\gamma_{c2} << \gamma$ the
barrier hight is essentially larger than $H_{\rm fluct}$.  The rough estimate
for this pseudocritical value is $\gamma_{c2} \sim 0.278$ at $\lambda=0.009$.

The fluctuations of $|\Phi|$ are around $\delta \phi \sim 0.6$ for all
considered values of $\gamma$ at $\lambda = 0.009, 0.0025, 0.001$, $\beta =
12$. It follows from our data (see also  Fig. \ref{fig.2_} ) that $\phi_m,
\langle |\phi|\rangle
>> \delta \phi$ at $\gamma_{c2} << \gamma$ while $\phi_m, \langle |\phi|\rangle \sim \delta \phi$ at
$\gamma_{c2} > \gamma$. Basing on these observations we expect that in the
region $\gamma_{c2} << \gamma$ the usual perturbation expansion around trivial
vacuum of spontaneously broken theory can be applied to the lattice Weinberg -
Salam model while in the FR $\gamma_c < \gamma < \gamma_{c2}$ it cannot be
applied. In the same way we define the pseudocritical value $\gamma_{c2}$ at
$\lambda = 0.001, 0.0025$. Namely, $\gamma_{c2} \sim 0.278$ for $\lambda =
0.009$; $\sim 0.262$ for $\lambda = 0.0025$; $\sim 0.258$ for $\lambda =
0.001$.

\section{The renormalized coupling}

In order to calculate the renormalized fine structure constant $\alpha_R = e^2/4\pi$
(where $e$ is the electric charge) we use the potential for infinitely heavy external
fermions.

We consider Wilson loops for the right-handed external leptons:
\begin{equation}
 {\cal W}^{\rm R}_{\rm lept}(l)  =
 \langle {\rm Re} \,\Pi_{(xy) \in l} e^{2i\theta_{xy}}\rangle.
\label{WR}
\end{equation}
Here $l$ denotes a closed contour on the lattice. We consider the following quantity
constructed from the rectangular Wilson loop of size $r\times t$:
\begin{equation}
 {\cal V}(r) = {\rm log}\, \lim_{t \rightarrow \infty}
 \frac{  {\cal W}(r\times t)}{{\cal W}(r\times (t+1))}.\label{vinf}
\end{equation}

\begin{figure}
\begin{center}
 \epsfig{figure=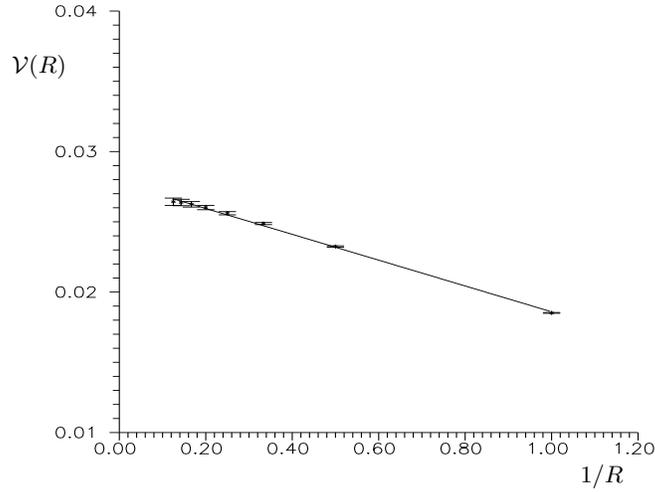,height=60mm,width=80mm,angle=0}
\end{center}
\begin{picture}(0,0)(0,0)
\put(-130,165){${\cal V}(R)$} \put(85,10){$\Large 1/R$}
\end{picture}
\begin{center}
\caption{\label{fig.1_1_} The potential ($T = 8$) for the right - handed
leptons vs. $1/R$ at $\gamma = 0.262$, $\lambda =0.0025$, and $\beta = 12$
(lattice $16^4$). }
\end{center}
\end{figure}

 Due to exchange by virtual photons at large enough
distances we expect the appearance of the Coulomb interaction
\begin{equation}
 {\cal V}(r) = -\frac{\alpha_R}{r} + const. \label{V1}
\end{equation}
It should be mentioned here, that in order to extract the renormalized value of $\alpha$
one may apply to $\cal V$ the fit obtained using the Coulomb interaction in momentum
space. The lattice Fourier transform then gives

\begin{eqnarray}
 {\cal V}(r) & = & -\alpha_R \, {\cal U}(r)+ const,\,
\nonumber\\
{\cal U}(r) & = & \frac{ \pi}{N^3}\sum_{\bar{p}\ne 0} \frac{e^{i p_3 r}}{{\rm sin}^2
p_1/2 + {\rm sin}^2 p_2/2 + {\rm sin}^2
 p_3/2}
 \label{V2}
\end{eqnarray}
Here $N$ is the lattice size, $p_i = \frac{2\pi}{L} k_i, k_i = 0, ..., L-1$. On large
enough lattices at $r << L$ both definitions approach each other. On the lattices we use
the values of the renormalized $\alpha_R$ extracted from (\ref{V1}) and (\ref{V2}) are
essentially different from each other. Any of the two ways, (\ref{V1}) or (\ref{V2}), may
be considered as the {\it definition} of the renormalized $\alpha$ on the finite lattice.
And there is no particular reason to prefer the potential defined using the lattice
Fourier transform of the Coulomb law in momentum space. Actually, our study shows that
the single $1/r$ fit approximates $\cal V$ much better. Moreover, the values of
renormalized $\alpha$ calculated using this fit are essentially closer to the tree level
estimate than that of calculated using the fit (\ref{V2}).

In practise instead of (\ref{vinf}) we use the potential that depends on additional
parameter $T$:
\begin{equation}
 {\cal V}(r,T) = {\rm log}\,
 \frac{  {\cal W}(r\times T)}{{\cal W}(r\times (T+1))}.
\end{equation}
For example, on the lattice $16^4$ the values  $T = 4,5,6,7,8$ are used; on the lattice
$12^3\times 16$ the values  $T = 4,5,6$ are used; on the lattice $8^3\times 16$ the value
$T = 4$ is used. As a result $\alpha_R = \alpha_R(T)$ may depend both on the lattice size
and on $T$. The dependence on $T$ was missed in \cite{Z2009} (where for lattices
$12^3\times 16, 16^4$ we used $T =5$,  while for the lattice $8^3\times 16$ we used
$T=4$).

\begin{figure}
\begin{center}
 \epsfig{figure=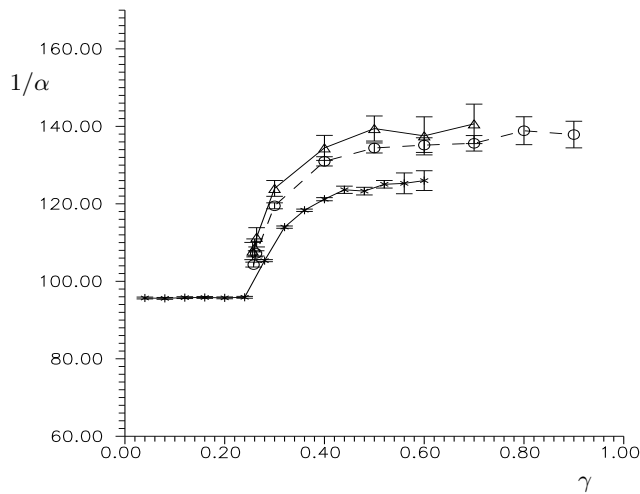,height=60mm,width=80mm,angle=0}
\end{center}
\begin{picture}(0,0)(0,0)
\put(-125,160){$1/\alpha$} \put(90,10){$\Large \gamma$}
\end{picture}
\begin{center}
\caption{\label{fig.1_} The inverse renormalized fine structure constant as a
function of $\gamma$ at $\lambda =0.0025$ and $\beta = 12$. It slowly
approaches the tree level estimate $\sim 150$ when $\gamma$ and the lattice
size are increased. Circles correspond to lattice $12^3\times 16$. Crosses
correspond to lattice $8^3\times 16$. Triangles correspond to lattice $16^4$.}
\end{center}
\end{figure}

On Fig. \ref{fig.1_1_} we represent as an example the dependence of the
potential for $T = 8$ on $1/R$. As it was already mentioned (\ref{V1})
approximates the potential much better than (\ref{V2}).   Therefore we used the
fit (\ref{V1}) to extract $\alpha_R$. This should be compared with the results
of \cite{14}, where for similar reasons the single $e^{-\mu r}/r$ fit (instead
of the lattice Yukawa fit) was used in order to determine the renormalized
coupling constant in the $SU(2)$ Gauge Higgs model.

Due to the dependence of $\alpha_R(T)$ on $T$  there is the essential
uncertainty in definition of $\alpha_R$ related to finite volume effects.  For
example, at $\gamma = 0.29$, $\lambda =0.009$, and $\beta = 12$ the value of
$\alpha_R$ calculated on the lattice $16^4$ varies between $\alpha_R(4) \sim
1/(93\pm 1)$ and $\alpha_R(8) \sim 1/(108\pm 2)$ (at the same time on the
lattice $8^3\times 16$ the value is $\alpha_R(4) = 1/(100\pm 1)$). At $\gamma =
0.274$, $\lambda =0.009$, and $\beta = 12$ the value of $\alpha_R$ calculated
on the lattice $20^3\times 24$ varies between $\alpha_R(4) \sim 1/(98\pm 1)$
and $\alpha_R(10) = 1/(106\pm 1)$ (at the same time on the lattice $8^3\times
16$ the value is $\alpha_R(4) = 1/(99\pm 1)$). Below for the lattice $8^3\times
16$ we use $T = 4$, for the lattice $12^3\times 16$ we use $T = 6$, for the
lattice $16^4$ we use $T = 8$. Therefore, the dependence on $T$ is absorbed
into the dependence on the lattice size. As an example, on Fig. \ref{fig.1_}
 we represent the renormalized fine structure constant
(calculated using the fit (\ref{V1})) at  $\lambda = 0.0025$, $ \beta = 12$.
The calculated values are to be compared with bare constant $\alpha_0 = 1/(4\pi
\beta)\sim 1/150$ at $\beta = 12$. One can see, that for $\gamma >>
\gamma_{c2}$ the tree level estimate is approached slowly while within the FR
the renormalized $\alpha$ differs essentially from the tree level estimate.
This is in correspondence with our supposition that the perturbation theory
cannot be valid within the FR while it works well far from the FR. The
dependence of $\alpha_R$ on the lattice size is clear: for the larger lattices
$\alpha_R$ approaches its tree level estimate faster than for the smaller ones.
Unfortunately, due to the difficulties  in simulation of the system at large
$\gamma$ we cannot observe this pattern in detail. At the present moment the
value of $\alpha_R$ most close to the tree level estimate
 is obtained on the lattice $12^3\times 16$ and is about $1/140$ (at
$\lambda = 0.0025, 0.001; \beta = 12; \gamma \sim 1$).

\section{Masses and the lattice spacing}

After fixing the unitary gauge $\Phi_1 \in R$, $\Phi_2 = 0$, $\Phi_1 \ge 0$ the
following variables are considered as creating a $Z$ boson and a $W$ boson,
respectively:

\begin{eqnarray}
  Z_{xy} & = & Z^{\mu}_{x} \;
 = - {\rm sin} \,[{\rm Arg} (U^{11}_{xy} e^{i\theta_{xy}}) ]
\nonumber\\
 W_{xy} & = & W^{\mu}_{x} \,= \,U_{xy}^{12} e^{-i\theta_{xy}}.\label{Z1}
\end{eqnarray}
Here, $\mu$ represents the direction $(xy)$. The electromagnetic $U(1)$
symmetry remains:
\begin{eqnarray}
 U_{xy} & \rightarrow & g^\dag_x U_{xy} g_y, \nonumber\\
 \theta_{xy} & \rightarrow & \theta_{xy} -  \alpha_y/2 + \alpha_x/2,
\end{eqnarray}
where $g_x = {\rm diag} (e^{i\alpha_x/2},e^{-i\alpha_x/2})$. There exists a $U(1)$
lattice gauge field, which is defined as
\begin{equation}
 A_{xy}  =  A^{\mu}_{x} \;
 = \,[-{\rm Arg} U_{xy}^{11} + \theta_{xy}]  \,{\rm mod} \,2\pi
\label{A}
\end{equation}
that transforms as $A_{xy}  \rightarrow  A_{xy} - \alpha_y + \alpha_x$. The field $W$
transforms as $W_{xy}  \rightarrow  W_{xy}e^{-i\alpha_x}$.

\begin{figure}
\begin{center}
 \epsfig{figure=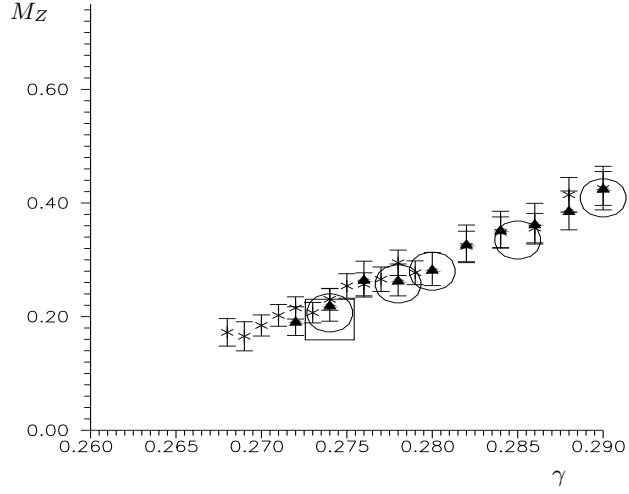,height=60mm,width=80mm,angle=0}
\end{center}
\begin{picture}(0,0)(0,0)
\put(-120,185){$M_Z$} \put(85,10){$\Large \gamma$}
\end{picture}
\begin{center}
\caption{\label{fig.3} Z - boson mass in lattice units at $\lambda =0.009$ and
$\beta = 12$ as a function of $\gamma$. Black triangles correspond to lattice
$12^3\times 16$. Crosses correspond to lattice $8^3\times 16$. Circles
correspond to lattice $16^4$. Square corresponds to lattice $20^3\times 24$.
(The error bars for lattices $16^3\times 16$ and $20^3\times 24$ are about of
the same size as the symbols used.)}
\end{center}
\end{figure}

 The $W$ boson field is charged with respect to the $U(1)$
symmetry. Therefore we fix the lattice Landau gauge in order to investigate the $W$ boson
propagator. The lattice Landau gauge is fixed via minimizing (with respect to the $U(1)$
gauge transformations) the following functional:
\begin{equation}
 F  =  \sum_{xy}(1 - \cos(A_{xy})).
\end{equation}
Then we extract the mass of the $W$ boson from the correlator
\begin{equation}
\frac{1}{N^6} \sum_{\bar{x},\bar{y}} \langle \sum_{\mu} W^{\mu}_{x}
(W^{\mu}_{y})^{\dagger} \rangle   \sim
  e^{-M_{W}|x_0-y_0|}+ e^{-M_{W}(L - |x_0-y_0|)}
\label{corW}
\end{equation}
Here the summation $\sum_{\bar{x},\bar{y}}$ is over the three ``space" components of the
four - vectors $x$ and $y$ while $x_0, y_0$ denote their ``time" components. $N$ is the
lattice length in "space" direction. $L$ is the lattice length in the "time" direction.

\begin{figure}
\begin{center}
 \epsfig{figure=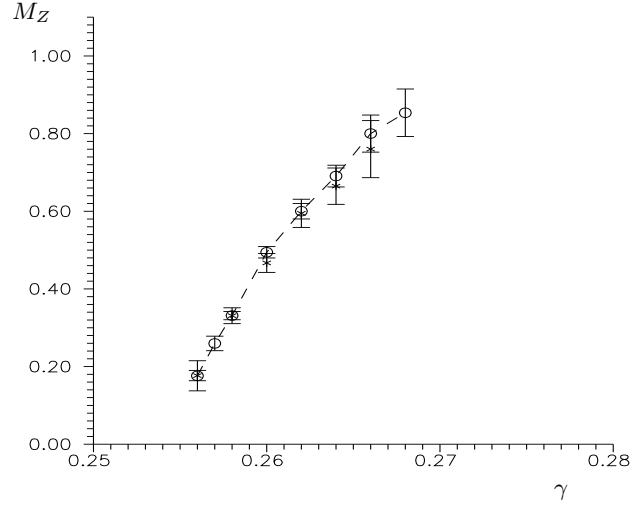,height=60mm,width=80mm,angle=0}
\end{center}
\begin{picture}(0,0)(0,0)
\put(-120,190){$M_Z$} \put(85,10){$\Large \gamma$}
\end{picture}
\begin{center}
\caption{\label{fig.3_3_} Z - boson mass in lattice units at $\lambda =0.001$
and $\beta = 12$. Circles correspond to lattice $8^3\times 16$. Crosses
correspond to  lattice $12^3\times 16$. }
\end{center}
\end{figure}

In order to evaluate the masses of the $Z$-boson and the Higgs boson we use the
correlators:
\begin{equation}
\frac{1}{N^6} \sum_{\bar{x},\bar{y}} \langle \sum_{\mu} Z^{\mu}_{x} Z^{\mu}_{y} \rangle
\sim
  e^{-M_{Z}|x_0-y_0|}+ e^{-M_{Z}(L - |x_0-y_0|)}
\label{corZ}
\end{equation}
and
\begin{equation}
  \frac{1}{N^6}\sum_{\bar{x},\bar{y}}(\langle H_{x} H_{y}\rangle - \langle H\rangle^2)
   \sim
  e^{-M_{H}|x_0-y_0|}+ e^{-M_{H}(L - |x_0-y_0|)},
\label{cor}
\end{equation}

In lattice calculations we used two different operators that create Higgs bosons: $ H_x =
|\Phi|$ and $H_x = \sum_{y} Z^2_{xy}$. In both cases $H_x$ is defined at the site $x$,
the sum $\sum_y$ is over its neighboring sites $y$.

\begin{figure}
\begin{center}
\epsfig{figure=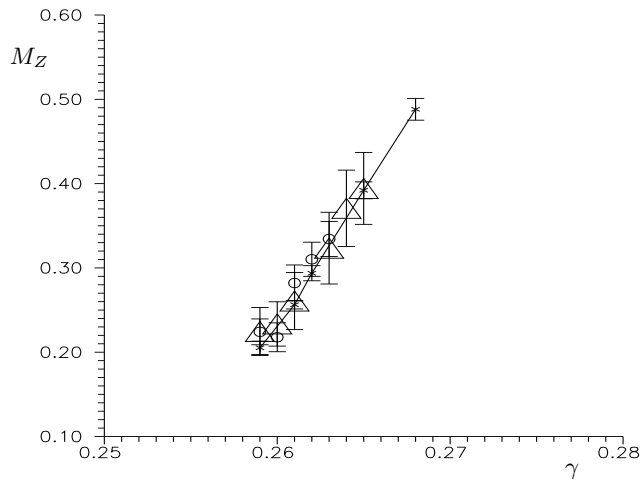,height=60mm,width=80mm,angle=0}
\end{center}
\begin{picture}(0,0)(0,0)
\put(-125,170){$M_Z$} \put(85,15){$\Large \gamma$}
\end{picture}
\begin{center}
\caption{\label{fig.3_2} Z - boson mass in lattice units at $\lambda =0.0025$
and $\beta = 12$. Circles correspond to lattice $12^3\times 16$. Crosses
correspond to lattice $8^3\times 16$. Triangles correspond to lattice $16^4$.}
\end{center}
\end{figure}

The physical scale is given in our lattice theory by the value of the $Z$-boson mass
$M^{phys}_Z \sim 91$ GeV. Therefore the lattice spacing is evaluated to be $a \sim [91
{\rm GeV}]^{-1} M_Z$, where $M_Z$ is the $Z$ boson mass in lattice units. The similar
calculations have been performed in \cite{VZ2008} for $\lambda = \infty$. It has been
found that the $W$ - boson mass contains an artificial dependence on the lattice size. We
suppose, that this dependence is due to the photon cloud surrounding the $W$ - boson. The
energy of this cloud is related to the renormalization of the fine structure constant.
 Therefore the $Z$ - boson mass was used
in order to fix the scale.

Our data show that $\Lambda= \frac{\pi}{a} = (\pi \times 91~{\rm GeV})/M_Z$ is
increased slowly with the decrease of $\gamma$ at any fixed $\lambda$. We
investigated carefully the vicinity of the transition point at fixed $\lambda =
0.001, 0.0025, 0.009$ and $\beta = 12$. It has been found that at the
transition point the value of $\Lambda$ is equal to $1.4 \pm 0.2$ TeV for
$\lambda = 0.009, 0.0025, 0.001$. Check of the dependence on the lattice size
($8^3\times 16$, $12^3\times 16$, $16^4$, $20^3\times 24$ at $\lambda =0.009$;
$8^3\times 16$, $12^3\times 16$, $16^4$ at $\lambda =0.0025$; $8^3\times 16$,
$12^3\times 16$ at $\lambda =0.001$) does not show an essential dependence of
this value on the lattice size. This is illustrated by Fig. \ref{fig.3},
Fig.\ref{fig.3_3_}, and Fig. \ref{fig.3_2}. From these figures it also follows
that at the value of $\gamma$ equal to $\gamma_{c2} (\sim 0.278$ for $\lambda =
0.009$; $\sim 0.262$ for $\lambda = 0.0025$; $\sim 0.258$ for $\lambda =
0.001$) the calculated value of the cutoff is  about $1$ TeV.

It is worth mentioning that the linear fit applied (in some vicinity of
$\gamma_c$) to the dependence of $M_Z$ on $\gamma$  predicts vanishing of
$M_Z(\gamma)$  at $\gamma$ equal to $\gamma_{c0} < \gamma_c$. Within the
statistical errors $\gamma_{c0} = 0.253\pm 0.001$ for $\lambda = 0.001$,
$\gamma_{c0} = 0.253\pm 0.001$ for $\lambda = 0.0025$, $\gamma_{c0} = 0.254\pm
0.001$ for $\lambda = 0.009$. We perform direct calculations within the region
$(\gamma_{c0}, \gamma_c)$ at $\lambda = 0.001, 0.0025$. These calculations show
that the fluctuations of the correlator (\ref{corZ}) are increased (compared
with the values of the correlator) fast when $\gamma$ is decreased. Already for
$\gamma = 0.255$ at $\lambda = 0.0025$ ($\gamma_c = 0.26$) and for $\gamma =
0.254$ at $\lambda = 0.001$ ($\gamma_c = 0.258$) the values of the correlator
at $|x_0 - y_0| > 0$ are smaller than the statistical errors.  Most likely, at
$\gamma \le \gamma_{c0}$ it is necessary to apply another gauge (like in pure
$SU(2)\times U(1)$ gauge model) in order to calculate gauge boson propagators.
At the present moment we do not estimate the scalar particle mass at
$\gamma_{c0}$ because of the lack of statistics. The behavior of the other
quantities is smooth at $\gamma \sim \gamma_{c0}$, no maximum of $\delta \phi$
or other susceptibilities is observed there (see, for example, Fig.
\ref{fig.6__}). Basing on our data it is natural to suppose that lattice gauge
boson mass may vanish at $\gamma \sim \gamma_{c0}$ although we do not observe
the correspondent pattern in details because of the strong fluctuations of
correlator (\ref{corZ}) near $\gamma_{c0}$. As it was mentioned above the
transition for the considered values of couplings is, most likely,  a
crossover. There are $3$ exceptional points: $\gamma_{c0}$, where lattice value
of $M_Z$ may vanish, $\gamma_c$, where scalar field condensate disappears, and
$\gamma_{c2}$ that denotes the boundary of the fluctuational region. This
situation is typical for the crossovers: different quantities change their
behavior at different points on the phase diagram. At the present moment we do
not exclude that the second order phase transition may take place at
$\gamma_{c0}$. This would happen if both mass parameters (Z boson mass and
scalar particle mass) vanish simultaneously at this point. The careful
investigation of the vicinity of $\gamma_{c0}$ is to be the subject of a
further research.

In the Higgs channel the situation is more difficult. Due to the lack of
statistics we cannot estimate the masses in this channel using the correlators
(\ref{cor}) at all considered values of coupling constants. Moreover, at
several points, where we have estimated the renormalized Higgs boson mass the
statistical errors are much larger than that of for the Z - boson mass. At the
present moment we can represent the data at four points on the lattice
$8^3\times16$: ($\gamma = 0.274$, $\lambda =0.009$, $\beta = 12$),  ($\gamma =
0.290$, $\lambda =0.009$, $\beta = 12$), ($\gamma = 0.261$, $\lambda =0.0025$,
$\beta = 12$), and ($\gamma = 0.257$, $\lambda =0.001$, $\beta = 12$).

The first point roughly corresponds to the position of the transition at
$\lambda =0.009$, $\beta = 12$ while the second point is situated deep within
the Higgs phase. These two points  correspond to bare Higgs mass around $270$
Gev.  At the point ($\gamma = 0.274$, $\lambda =0.009$, $\beta = 12$) we have
collected enough statistics to calculate correlator (\ref{cor}) up to the
"time" separation $|x_0-y_0| = 4$. The value $\gamma = 0.274$ corresponds
roughly to the position of the phase transition. We estimate at this point $M_H
= 300 \pm 40$ Gev.  At the point ($\gamma = 0.29$, $\lambda =0.009$, $\beta =
12$) we calculate the correlator with reasonable accuracy up to $|x_0-y_0| =
3$.  At this point $M_H = 265 \pm 70$ Gev.

For  $\lambda = 0.001, 0.0025$ we calculate the Higgs boson mass close to the
transition points. Similar to the case $\lambda = 0.009$ we do not observe here
essential deviation from the tree level estimates. Namely, for $\lambda =
0.001, \gamma = 0.257$
 we have $M_H = 90 \pm 20$ GeV (tree level value is $M^0_H \sim 100$ GeV). In this point
  we have collected enough statistics to calculate
correlator (\ref{cor}) up to the "time" separation $|x_0-y_0| = 8$. For $\lambda =
0.0025, \gamma = 0.261$
 we have $M_H = 170 \pm 30$ GeV (tree level value is $M^0_H \sim 150$ GeV). In this point
  we have collected enough statistics to calculate
correlator (\ref{cor}) up to the "time" separation $|x_0-y_0| = 4$. It is worth
mentioning that in order to calculate $Z$ - boson mass we fit correlator
(\ref{corZ}) for $8 \ge |x_0-y_0| \ge 1$.

\section{Nambu monopole density }

The worldlines of the quantum Nambu monopoles can be extracted from the field
configurations according to Eq. (\ref{Am}).
The monopole density is defined as $ \rho = \left\langle \frac{\sum_{\rm links}|j_{\rm
link}|}{4V^L}
 \right\rangle,$
where $V^L$ is the lattice volume.

\begin{figure}
\begin{center}
 \epsfig{figure=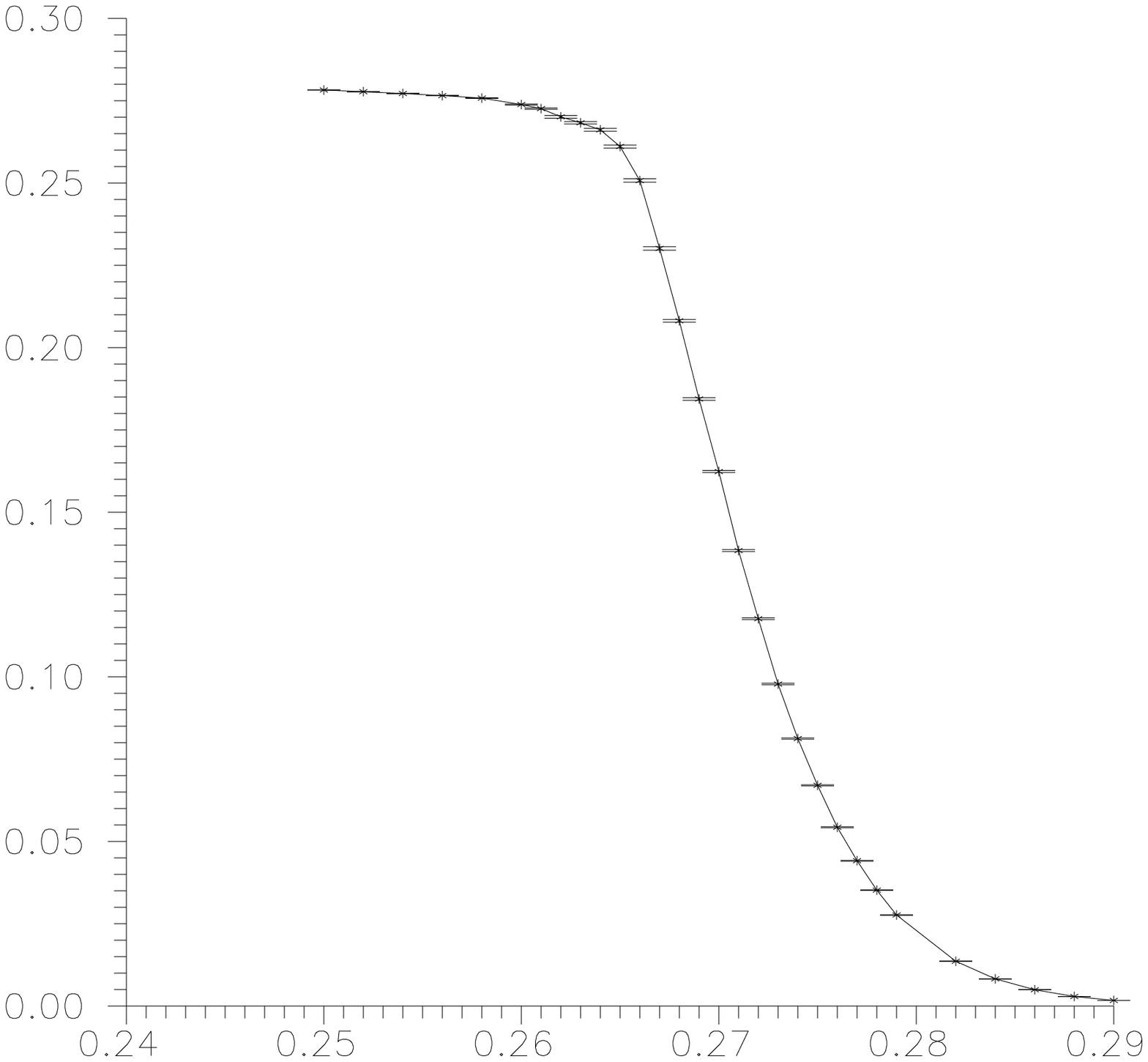,height=60mm,width=80mm,angle=0}
\end{center}
\begin{picture}(0,0)(0,0)
\put(-125,175){$\rho$} \put(85,10){$\Large \gamma$}
\end{picture}
\begin{center} \caption{\label{fig.5_} Nambu monopole density as a function of
$\gamma$ at $\lambda =0.009$ and $\beta = 12$. (Lattice $8^3\times16$.)  }
\end{center}
\end{figure}

\begin{figure}
\begin{center}
 \epsfig{figure=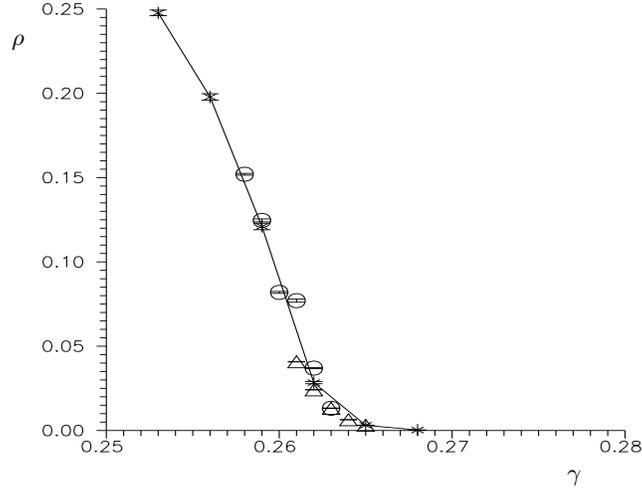,height=60mm,width=80mm,angle=0}
\end{center}
\begin{picture}(0,0)(0,0)
\put(-125,175){$\rho$} \put(85,10){$\Large \gamma$}
\end{picture}
\begin{center}\caption{\label{fig.5_1} Nambu monopole density as a function of $\gamma$ at
$\lambda =0.0025$ and $\beta = 12$. Circles correspond to lattice $12^3\times
16$. Crosses correspond to  lattice $8^3\times16$. Triangles correspond to
lattice $16^4$. }
\end{center}
\end{figure}

\begin{figure}
\begin{center}
 \epsfig{figure=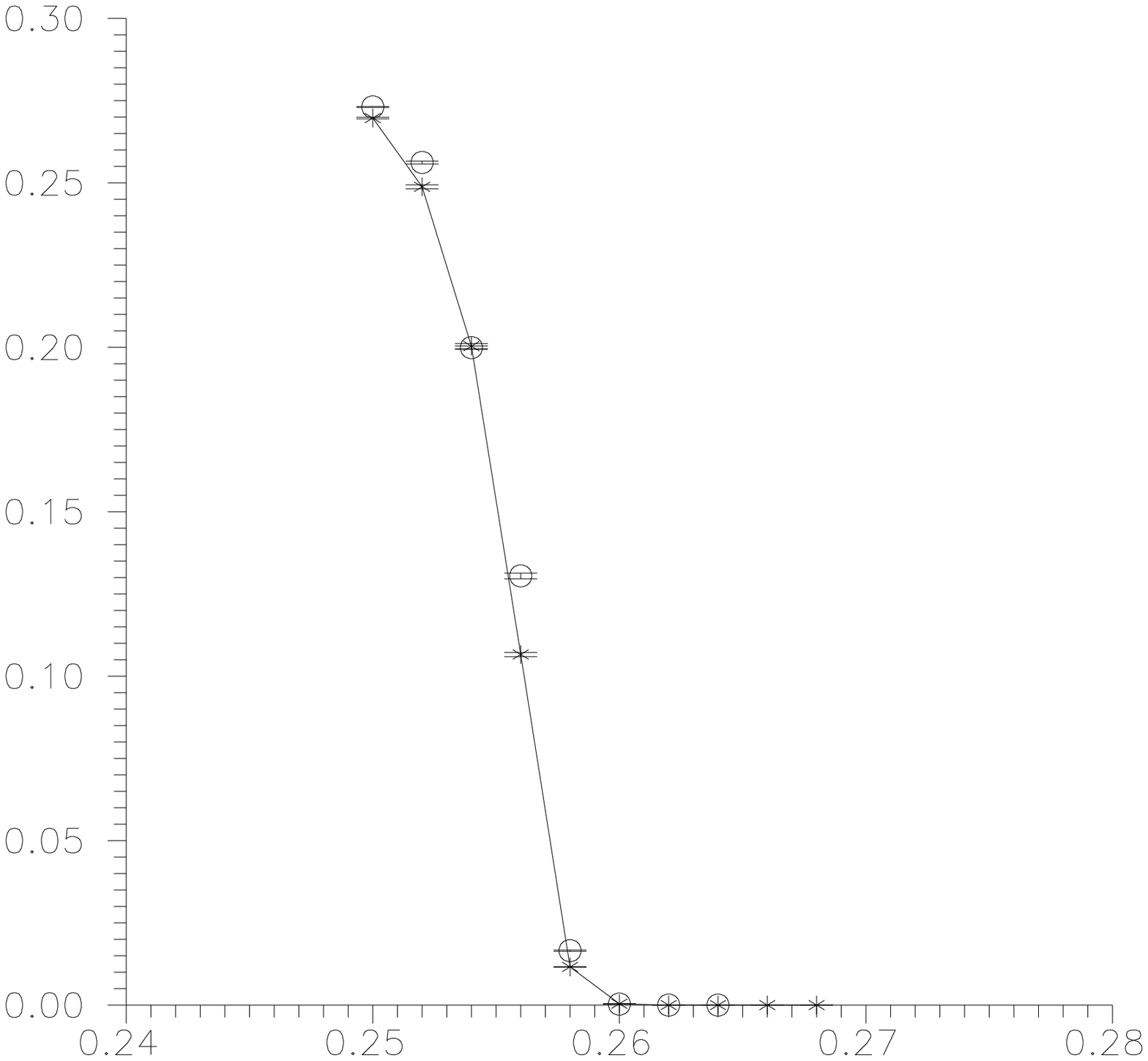,height=60mm,width=80mm,angle=0}
\end{center}
\begin{picture}(0,0)(0,0)
\put(-125,175){$\rho$} \put(85,10){$\Large \gamma$}
\end{picture}
\begin{center}\caption{\label{fig.5_1_} Nambu monopole density as a function of $\gamma$ at
$\lambda =0.001$ and $\beta = 12$. Circles correspond to lattice $12^3\times
16$. Crosses correspond to  lattice $8^3\times16$. }
\end{center}
\end{figure}

On Fig \ref{fig.5_}, Fig. \ref{fig.5_1}, and Fig. \ref{fig.5_1_}  we represent
Nambu monopole density as a function of $\gamma$ at $\lambda = 0.009, 0.0025,
0.001$, $\beta = 12$. The value of monopole density at $\gamma_c$ is around
$0.1$.

According to the classical picture the Nambu monopole size is of the order of $M^{-1}_H$.
Therefore, for example, for $a^{-1} \sim 430$ Gev and $M_H \sim 300, 150, 100$ Gev the
expected size of the monopole is about a lattice spacing. The monopole density around
$0.1$ means that among $10$ sites there exist $4$ sites that are occupied by the
monopole. Average distance between the two monopoles is, therefore, less than $1$ lattice
spacing and it is not possible at all to speak of the given configurations as of
representing the physical Nambu monopole.

At $\gamma = \gamma_{c2}$ the Nambu monopole density is of the order of $0.01$. This
means that among about $25$ sites there exists one site that is occupied by the monopole.
Average distance between the two monopoles is, therefore, between one and two lattice
spacings. We see that at this value of $\gamma$ the average distance between Nambu
monopoles is of the order of their size.

We summarize the above observations as follows. Within the fluctuational region the
configurations under consideration do not represent single Nambu monopoles. Instead these
configurations can be considered as the collection of monopole - like objects that is so
dense that the average distance between the objects is of the order of their size. On the
other hand, at $\gamma
>> \gamma_{c2}$ the considered configurations do represent single Nambu
monopoles and the average distance between them is much larger than their size. In other
words out of the FR vacuum can be treated as a gas of Nambu monopoles while within the FR
vacuum can be treated as a liquid composed of monopole - like objects.

It is worth mentioning that somewhere inside the $Z$ string connecting the classical
Nambu monopoles the Higgs field is zero: $|\Phi| = 0$. This means that the $Z$ string
with the Nambu monopoles at its ends can be considered as an embryo of the symmetric
phase within the Higgs phase. We observe that the density of these embryos is increased
when the phase transition is approached. Within the fluctuational region the two phases
are mixed, which is related to the large value of Nambu monopole density.

That's why we come to the conclusion that vacuum of lattice Weinberg - Salam model within
the FR has nothing to do with the continuum perturbation theory. This means that the
usual perturbation expansion around trivial vacuum (gauge field equal to zero, the scalar
field equal to $(\phi_m,0)^T$) cannot be valid within the FR.  This might explain why we
do not observe in our numerical simulations the large values of $\Lambda$ predicted by
the conventional perturbation theory.

\section{Conclusions}

In the present paper we demonstrate that while approaching continuum physics in
lattice Weinberg - Salam model one encounters the nonperturbative effects.
Namely, the continuum physics is to be approached in the vicinity of the
transition between the physical Higgs phase and the  symmetric phase of the
model (in the symmetric phase the scalar field is not condensed). The
ultraviolet cutoff is increased when the transition point is approached along
the line of constant physics. There exists the fluctuational region (FR) on the
phase diagram of the lattice Weinberg - Salam model. This region is situated in
the vicinity of the transition between the Higgs phase and the symmetric phase
(where scalar field is not condensed). According to our data this transition
is, most likely,  a crossover. We localize its position at the point
$\gamma_c(\lambda, \beta, \theta_W)$, where the
 scalar field condensate disappears. We calculate the effective constraint
potential $V(\phi)$ for the Higgs field. It has a minimum at the nonzero value
$\phi_m$ in the physical Higgs phase. At the considered values of $\lambda,
\beta, \theta_W$ for $\gamma$ between $\gamma_c$ and $\gamma_{c2}$
($\gamma_{c2}$ is in the Higgs phase) the fluctuations of the scalar field
become of the order of $\phi_m$. Moreover, the "barrier hight" $H = V(0) -
V(\phi_m)$ is of the order of $V(\phi_m + \delta \phi)- V(\phi_m)$, where
$\delta \phi$ is the fluctuation of $|\Phi|$. Therefore, we refer to this
region as to FR.

The scalar field must be equal to zero somewhere within the classical Nambu monopole.
That's why this object can be considered as an embryo of the unphysical symmetric phase
within the physical Higgs phase of the model. We investigate properties of the quantum
Nambu monopoles. Within the FR they are so dense that the average distance between them
becomes of the order of their size. This means that the two phases are mixed within the
FR. All these results show that the vacuum of lattice Weinberg - Salam model in the FR is
essentially different from the trivial vacuum used in the conventional perturbation
theory. As a result the use of the perturbation theory in this region is limited.

 Our numerical results show that at $M_H$ around $270, 150, 100$ GeV and the bare fine structure
constant around $1/150$ the maximal value of the cutoff admitted out of the FR
for the considered lattice sizes cannot exceed the value around $1$ Tev. Within
the FR the larger values of the cutoff can be achieved in principle. The
maximum for the value of the cutoff $\Lambda_c$ within the Higgs "phase" of the
lattice model is achieved at the point of the transition to the region of the
phase diagram, where the scalar field is not condensed. Our estimate for this
value is $\Lambda_c = 1.4 \pm 0.2$ Tev for the considered lattice sizes. Far
from the fluctuational region the behavior of the lattice model in general is
close to what we expect basing on the continuous perturbation theory. As it was
already mentioned at the considered values of couplings the transition is, most
likely, a crossover. This follows from the observation that various quantities
(Z boson mass, the fluctuation of the scalar field etc) do not depend on the
lattice size at the transition point. Within the symmetric "phase" of the
lattice model (where the scalar field is not condensed) in some vicinity of the
transition between this phase and the Higgs phase (where the scalar field is
condensed) the lattice gauge boson masses do not vanish. The statistical error
for $M_Z$ is increased fast when $\gamma$ is decreased starting from the
pseudocritical value $\gamma_c$. At $\gamma \le \gamma_{c0} < \gamma_c$ (within
the symmetric phase) the values of the $Z$ - boson correlator (\ref{corZ}) are
smaller than the statistical errors. Therefore, our procedure cannot give the
values of gauge boson masses in this region. Most likely, here the other gauge
is to be applied in order to calculate gauge boson propagators (we used in our
simulations the Unitary gauge). It is worth mentioning that the perturbation
theory predicts zero gauge boson masses within the symmetric phase. Most
likely, this prediction is failed within the interval $(\gamma_{c0}, \gamma_c)$
due to nonperturbative effects.

An important question is how to treat finite volume effects that are present in
all observables that contain long - ranged Electromagnetic Coulomb
interactions. In particular, we see that these effects are strong in
renormalized fine structure constant (about $10\%$  when the lattice size
varies from $8^3\times 16$ to $16^4$) and in the mass of electrically charged
$W$ - boson. On the other hand all observables related to $SU(2)$ constituent
of the model do not possess essential dependence on the lattice size. In
particular, $Z$ - boson mass $M_Z$ (and the cutoff $\Lambda$), density
$\rho_{\rm Nambu}$ of Nambu monopoles \footnote{Nambu monopoles in practise
correspond to $SU(2)$ variables as the monopole configurations extracted from
the Hypercharge U(1) field disappear at realistic values of coupling
constants.}, fluctuation of the scalar field $\delta \phi$ as well as the
position of the transition between the "phases" of the lattice model
practically do not depend on the lattice size. Our point of view is that the
influence of long - ranged Electromagnetic interactions on these observables is
negligible compared to their tree - level and nonperturbative constituents.
Actually, Electromagnetic interactions can be taken into account
perturbatively, with the renormalized $\alpha \sim 1/100$ as the parameter of
the perturbation expansion. This was the reason why in the previous numerical
studies of $SU(2)$ Gauge - Higgs model the $U(1)$ constituent of Weinberg -
Salam model was completely disregarded \cite{1,2,3,4,5,6,7,8,9,10,11,12,13,14}.
To summarize, we suppose that in spite of the presence of finite volume effects
in fine structure constant and $W$ boson mass, the calculated values of $M_Z$ ,
$\Lambda$, $\rho_{\rm Nambu}$, $\delta \phi$ etc can be considered as free of
these effects\footnote{The inverse seem to us incorrect: influence of
nonperturbative effects on $\alpha_R$ is not suppressed by any small factor. We
indeed observe that in the FR, where nonperturbative effects are large the
renormalized $\alpha$ differs from its bare value by about $50\%$ while far
from the FR the difference is within $10\%$ (for the lattice size
$12^3\times16$).} (up to the perturbations suppressed by the factor $\alpha
\sim 1/100$).

Basing on our data it is natural to suppose that lattice gauge boson mass may
vanish at $\gamma \sim \gamma_{c0}$ although we do not observe the
correspondent pattern in details because of the strong fluctuations of
correlator (\ref{corZ}) near $\gamma_{c0}$. If so, there exist $3$
pseudocritical points: $\gamma_{c0}$, where lattice value of $M_Z$ vanishes (at
this point the cutoff calculated as $\Lambda= (\pi \times 91~{\rm GeV})/M_Z$
tends to infinity), $\gamma_c$, where scalar field condensate disappears, and
$\gamma_{c2}$ that denotes the boundary of the fluctuational region (at $\gamma
\sim \gamma_{c2}$ the average distance between Nambu monopoles becomes of the
order of their size). This situation is typical for the crossovers: different
quantities change their behavior at different points on the phase diagram.
There still exists the possibility that the point $\gamma_{c0}$ corresponds to
the second order phase transition (this may happen if, in addition, the scalar
particle mass vanishes at $\gamma_{c0}$). However, the absence of a peak in the
scalar field fluctuation and in susceptibility (\ref{chiH}) at this point
indicates that this is a crossover. Actually, this possibility is to be checked
carefully but this is to be a subject of another work. There is an important
question: what is the relation between the conventional Electroweak physics and
the regions $(\gamma_{c0}, \gamma_c)$ and $(\gamma_{c}, \gamma_{c2})$. Our
expectation is that both these regions have nothing to do with real continuum
physics. For the first region this is more or less obvious: there the scalar
field is not condensed that contradicts with the usual spontaneous breakdown
pattern. As for the second region, the situation is not so obvious. However,
there the nonperturbative effects are strong and the Nambu monopoles dominate
vacuum that seems to us unphysical. With all mentioned above we come to the
conclusion that our data indicate the appearance of the maximal value of the
cutoff in Electroweak theory that cannot exceed the value of the order of $1$
TeV. This prediction is made basing on the numerical investigation of the
lattice model on the finite lattices. However, as it was mentioned above, our
main results do not depend on the lattice size.

\begin{acknowledgments}

 This work was partly supported by RFBR grants 09-02-00338, 08-02-00661, by
Grant for leading scientific schools 679.2008.2. The numerical simulations have
been performed using the facilities of Moscow Joint Supercomputer Center.

\end{acknowledgments}

\end{document}